\documentclass[11pt,twoside]{article}

\usepackage{asp2006}
\usepackage{epsf}
\usepackage{psfig}
\usepackage{lscape}
\usepackage{natbib,graphicx}

\begin{document}
\title{Smoothed Particle Hydrodynamics: Turbulence and MHD}   %%% Fill in title
\author{Daniel J. Price}   %%% Fill in author names
\affil{Centre for Stellar and Planetary Astrophysics, School of Mathematical Sciences, Monash University, Clayton Vic 3800, Australia}

\author{Christoph Federrath}
\affil{Zentrum f\"ur Astronomie der Universit\"at Heidelberg, Institut f\"ur Theoretische Astrophysik, Albert-Ueberle-Str.~2, D-69120 Heidelberg, Germany}
 %%% Fill in author affiliations

\begin{abstract} %%% Abstract to run on from here.
 In this paper we discuss recent applications of the Smoothed Particle Hydrodynamics (SPH) method to the simulation of supersonic turbulence in the interstellar medium, as well as giving an update on recent algorithmic developments in solving the equations of magnetohydrodynamics (MHD) in SPH. Using high resolution calculations (up to 134 million particles), we find excellent agreement with grid-based results on a range of measures including the power spectrum slope in both the velocity field and the density-weighted velocity $\rho^{1/3} v$, the latter showing a Kolmogorov-like $k^{-5/3}$ scaling as proposed by Kritsuk et al. (2007). We also find good agreement on the statistics of the Probability Distribution Function (PDF) and structure functions, independently confirming the scaling found by \citet*{sfk08}. On Smoothed Particle Magnetohydrodynamics (SPMHD) we have recently wasted a great deal of time and effort investigating the vector potential as an alternative to the Euler potentials formulation, in the end concluding that using the vector potential has even more severe problems than the standard (${\bf B}$-field based) SPMHD approach.
\end{abstract}

%%% MAIN BODY OF TEXT GOES HERE. CONSULT "INSTRUCTIONS FOR AUTHORS USING
%%% LATEX2E MARKUP", SECTIONS 2.3-2.6 FOR HELP WITH EQUATIONS, FIGURES,
%%% AND TABLES.

\section{Introduction}
 Turbulence and magnetic fields are thought to be two of the most important ingredients in the star formation process, so much so that there remains ongoing debate --- both observational and theoretical --- as to which one is \emph{the} controlling factor \citep[e.g.][]{mk04,km09}. We have only recently begun to understand the role of either in depth, primarily as a result of our increased ability to simulate both in numerical calculations of the star formation process.

 %What we do know about turbulence in the interstellar medium (ISM) is that it is ubiquitously observed to be highly supersonic (Mach$\sim$5-20) which places the dynamics of molecular clouds far from the incompressible regime in which turbulence is studied in the laboratory. Thus it is unclear that any of the phenomenological models developed to explain turbulence on earth (e.g. \citealt{kolmogorov41,sl94}) can be used to understand the ISM. What we know from observations is that molecular clouds show a clear velocity dispersion-size relation over a wide range of scales suggestive of a turbulent cascade \citep{larson81,hb04}, and that the the mass distribution of dense ``cores'' (an ill-defined term related to ill-defined dense subregions of molecular clouds) roughly follows a log-normal distribution \citep{alvesetal07}, similar to (possibly even directly related to) the mass distribution of newborn stars --- the Initial Mass Function (IMF) \citep{muenchetal02,chabrier03}.

Several authors have proposed that turbulence in the interstellar medium (ISM) can be used to predict the form of the stellar initial mass function (IMF)  thus providing a `theory' of star formation \citep{pn02,HennebelleChabrier2008}. Any such theory requires consensus on the basic statistical characteristics of turbulence in the supersonic regime and assumes that these are universal --- that is, independent of boundary conditions and driving mechanisms --- for which there is some observational support \citep[e.g.][]{hb04}. However there exists considerable disagreement between results obtained using different numerical codes, most recently between \citet{padoanetal07} who claim that the statistics of turbulence are universal and \citet{bpetal06} who claim that they are not, based on calculations utilising both a Smoothed Particle Hydrodynamics (SPH) code and a grid-based Total-Variation-Diminishing (TVD) code.

 This kind of disagreement, and the need within star formation theory for a consensus on the basic statistics of supersonic turbulence has prompted at least two major code comparison projects over the last few years, the ``Potsdam'' comparison \citep{kitsionasetal09}, comparing decaying, hydrodynamic turbulence and the KITP comparison (unpublished), comparing decaying hydrodynamic and magnetohydrodynamic (MHD) turbulence. However these comparisons suffer from the limited time evolution that can be obtained from a decaying turbulence simulation as well as the numerical issues posed by starting from an evolved snapshot produced by a particular code.
 
  We have therefore undertaken our own very detailed comparison of \emph{driven} turbulence using just two codes, an SPH code, \textsc{phantom}, and a grid code, \textsc{flash} (used in uniform grid mode) taken to be broadly representative of their class of codes, in order to see whether agreement can be reached (and at what resolution) on the statistics of supersonic turbulence appropriate to the ISM. The results are published in full in \citet{pf10} and we only provide a snapshot here of the main results (\S\ref{sec:turbulence}), referring the reader to that paper for more detailed information.

What we are not yet able to address is the combination of \emph{driven} turbulence and MHD in SPH, primarily because of the limitations posed by the Euler potentials formulation used as the only method that sufficiently maintains the divergence-free ($\nabla\cdot{\bf B} = 0$) constraint on the magnetic field such that star formation calculations can be performed \citep[e.g.][]{pb07,pb08,pb09} without the stars themselves exploding due to numerical errors (see \S\ref{sec:mhd}). Following a suggestion from Axel Brandenburg, we therefore embarked on a quest to examine whether or not the use of the vector potential could similarly solve the divergence problem in the context of Smoothed Particle Magnetohydrodynamics (SPMHD) without the associated topological restrictions of the Euler potentials \citep[see][]{brandenburg10}. In short, this turned out not to be the case \citep[read the horror that is][]{price10}. However, for the pleasure of the reader we describe the tortuous journey in \S\ref{sec:mhd}

\section{Turbulence: Grids vs. SPH}
\label{sec:turbulence}
 We adopted the usual approach to the artificial production of turbulence in numerical codes, using periodic boundary conditions and applying a ``large-scale'' driving in fourier space with random amplitudes and phases that are slowly changed over a timescale roughly comparable to the box crossing time (see \citealt{federrathetal08} for details). In order to obtain as close a comparison as possible we pre-generated the random driving pattern for all times during the simulation such that both codes use an identical forcing. Studying driven turbulence means that there are no numerical issues in either code relating to the (very simple) uniform initial conditions.
\begin{figure}
\begin{center}
\includegraphics[angle=270,width=\textwidth]{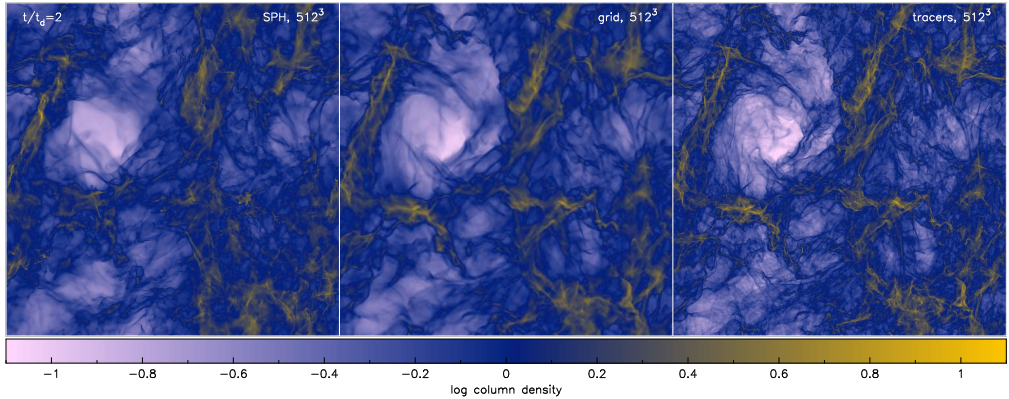}
\caption{Column density in Mach~10 turbulence calculations employing $512^{3}$ computational elements using the SPH code (\textsc{phantom}, left), using the grid code (\textsc{flash}, middle), and using an SPH density calculated from the \textsc{flash} tracer particles (right).}
\label{fig:coldens}
\end{center}
\end{figure}

\begin{figure}
\begin{center}
\includegraphics[angle=270,width=0.65\textwidth]{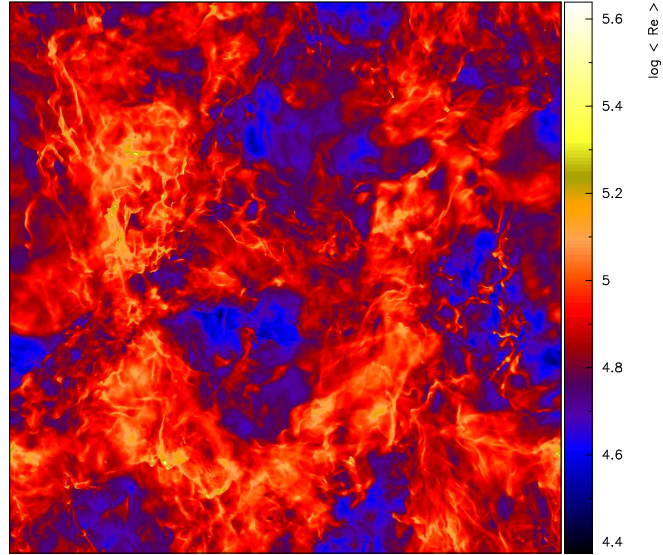}
\caption{Line-of-sight averaged Reynolds number $\int (\rho \mathrm{Re}) \mathrm{dz}/ \int \rho  \mathrm{dz}$ in the 512$^{3}$ SPH Mach~10 turbulence calculation. The mean Reynolds number at this resolution is around $10^{5}$, with individual Reynolds numbers reaching $\sim 10^{6}$ in the densest regions.}
\label{fig:reynolds}
\end{center}
\end{figure}

 Figure~\ref{fig:coldens} shows the column density after two dynamical times (adopting the usual definition $t_{d} = L/(2\mathcal{M})$ where $L$ is the box size and $\mathcal{M}$ is the RMS Mach number) from Mach~10 calculations employing $512^{3}$ computational elements in both grid cells and SPH particles. Whilst $512^{3}$ is only a moderate resolution in current grid-based turbulence simulations, our use of $512^{3}$ or approximately 134 million SPH particles makes this easily the highest resolution turbulence calculation ever performed with SPH. Indeed, given that SPH is a much more expensive computational method compared to uniform-grid calculations \citep{kitsionasetal09} it is reassuring that dense regions appear significantly better resolved in the SPH results (with adaptive smoothing lengths the effective SPH resolution at any point is $\sim n^{1/3}/L$ where $n$ is the local particle density). Since any viscosity in SPH must be explicitly added to the calculation (in the form of artificial viscosity, AV), the Reynolds number at any point in the flow can also be explicitly calculated. Figure~\ref{fig:reynolds} shows the line-of-sight averaged Reynolds number, $\langle \mathrm{Re} \rangle \equiv \int (\rho \mathrm{Re}) {\rm dz}/ \int \rho \phantom{.}{\rm dz}$, where the Reynolds number for each particle $a$ has been calculated according to $\mathrm{Re} = \vert v \vert L /\nu$, where $\vert v\vert$ is the magnitude of the particle velocity, $L$ is the box size and the viscosity $\nu$ is given by $\nu = 1/10 \alpha_{AV} c_{s} h$ in 3D \citep[e.g.][]{murray96,lp10}. The mean $\mathrm{Re}$ is around $10^{5}$ in the highest resolution calculation, with individual $\mathrm{Re}$'s reaching $\sim 10^{6}$ in the densest regions.

\begin{figure}
\begin{center}
\includegraphics[angle=270,width=0.75\textwidth]{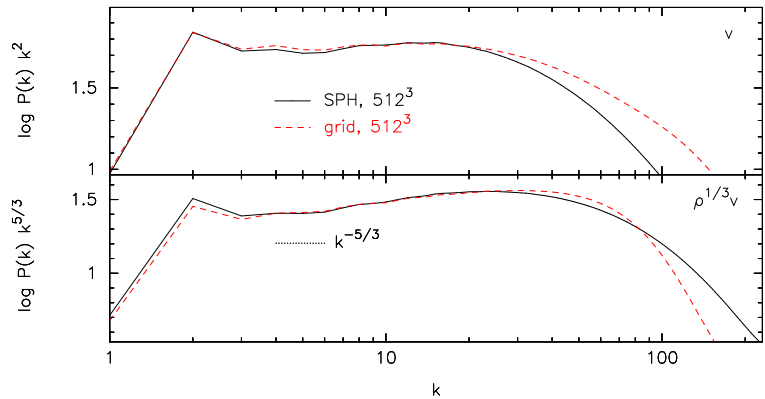}
\caption{Compensated, time averaged power spectra in the $512^{3}$ grid (dashed, red) and SPH (solid, black) turbulence calculations, showing the velocity field (top, compensated by $k^{2}$) and for the quantity $\rho^{1/3} v$ (bottom, compensated by $k^{5/3}$) which appears to scale in a Kolmogorov-like way.}
\label{fig:pspec}
\end{center}
\end{figure}
 
 Another unique aspect  of our comparison was adding Lagrangian tracer particles to the \textsc{flash} calculation, where we have used the SPH density calculation routine from \textsc{phantom} in order to calculate a density field for the tracer particles distribution (right panel of Figure~\ref{fig:coldens}). Doing so reveals an amazing resolution of sub-grid structures compared to the raw grid-based density field (middle panel). The caveat is that in the Probability Distribution Function it is clear that at least some of this structure may be an artefact of the tracer particle dynamics below the grid scale, in that the density PDF on the tracers in the low resolution runs does not appear to match the grid-based density PDF at higher resolution.
 
 Whilst the SPH is generally better resolved in the density field, reflected also in the PDFs \citep[see][]{pf10}, it is not necessarily true for more volumetric statistics, such as power spectra (Figure~\ref{fig:pspec}) where we find comparable results roughly when the number of computational elements is equal (Figure~\ref{fig:pspec}). The power spectra of the density-weighted quantity $\rho^{1/3} v$ show a Kolmogorov-like $k^{-5/3}$ scaling at high resolution, in agreement with the `new universality' proposed by \citet{kritsuketal07}. SPH appears more dissipative in pure velocity statistics (top panel of Figure~\ref{fig:pspec}) but correspondingly less dissipative in density-weighted statistics (bottom panel of Figure~\ref{fig:pspec}).

 Overall, we find very good agreement between the SPH and grid results on the statistics of supersonic turbulence, provided sufficient resolution is employed in either code, where ``sufficient'' depends on the desired statistic --- we require roughly $512^{3}$ computational elements in either code in order to resolve the inertial range in the power spectrum, whereas a good estimate of the PDF is obtained at lower resolutions. We find excellent agreement at high resolution between our measured structure function scaling and results obtained recently by \citet*{sfk08} at Mach~6.

\section{Smoothed Particle Magnetohydrodynamics: an update}
\label{sec:mhd}
 Whilst SPH is widely used for solving the equations of hydrodynamics in many astrophysical applications, the incorporation of magnetic fields into the SPH algorithm has a long and somewhat tortuous history. This is primarily due to a numerical instability that occurs if one tries to formulate the SPMHD equations such that momentum is conserved exactly (i.e., as the gradient of a stress tensor). In particular, SPH in its most basic form relies on a net repulsive force between particles to preserve order in the particle distribution. In MHD the stress tensor can become negative when the magnetic pressure is greater than the gas pressure so the particles can attract each other unstoppably. However physically any such force along the line of sight is spurious, related to the non-zero divergence of the magnetic field. Amongst other things (e.g. formulating shock-capturing dissipative terms, incorporating spatially variable smoothing lengths), dealing with this ``clumping instability'' was one of the key steps to a robust algorithm developed in \citet{pm04a,pm04b,pm05}.
\begin{figure}
\begin{center}
\includegraphics[width=0.45\textwidth]{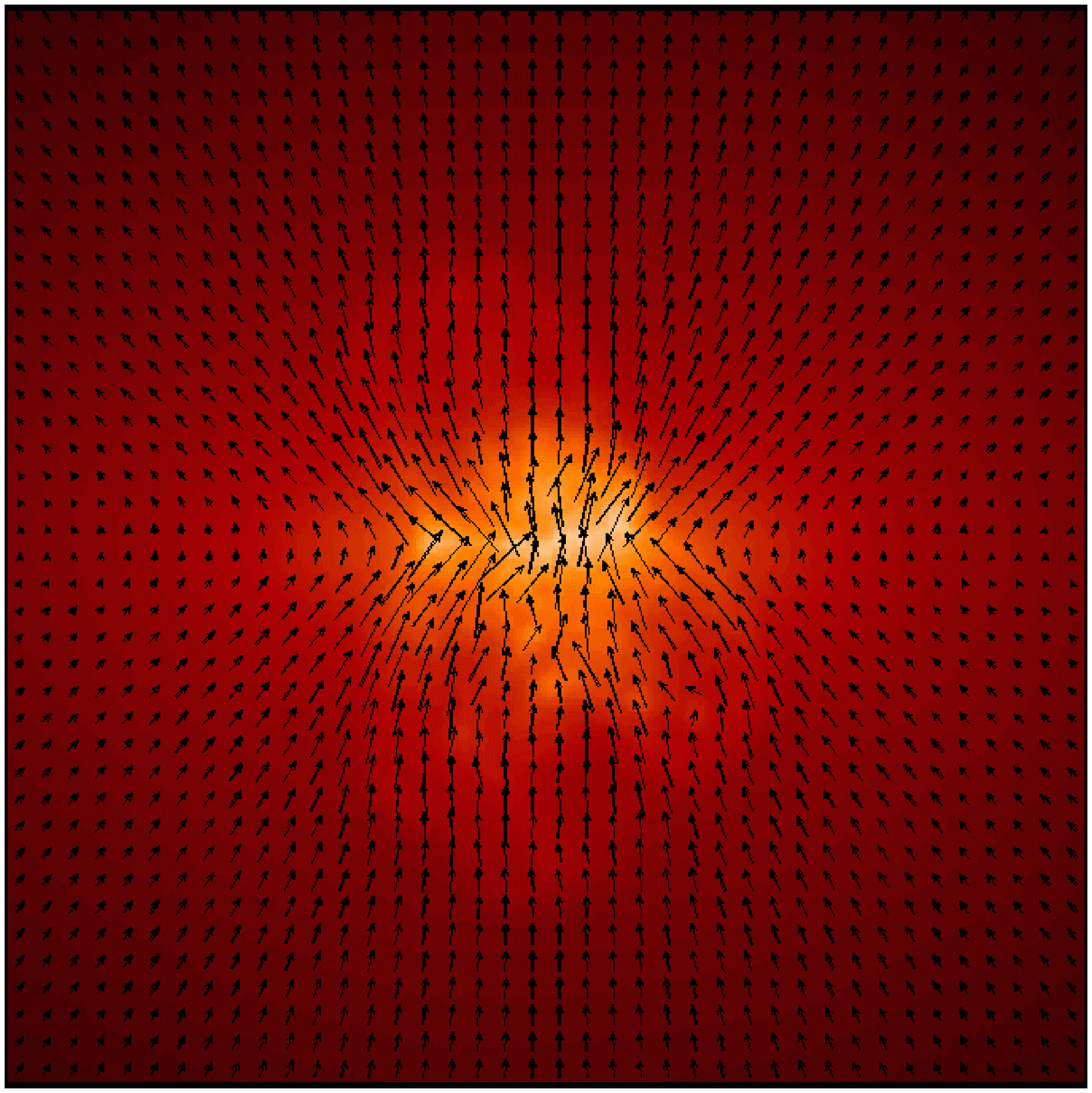}
\includegraphics[width=0.45\textwidth]{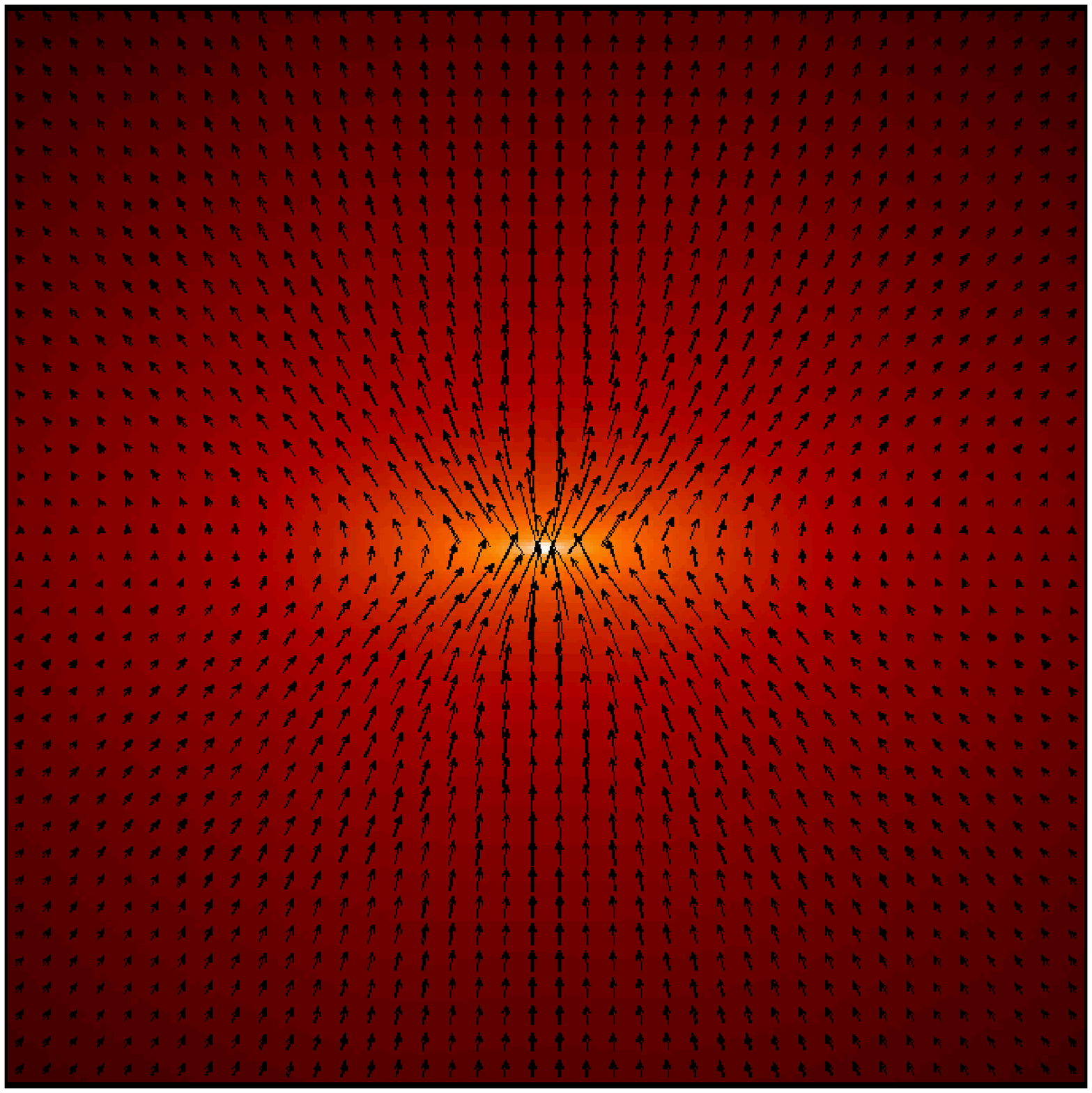}
\caption{The meaning of ``maintaining the divergence constraint'' in Smoothed Particle Magnetohydrodynamics. In the absence of effective control of divergence errors (left) the collapsing protostar explodes due to large numerical errors in the force in the region of extreme magnetic field gradients. Using the Euler potentials (right) these errors can be effectively controlled (and simulations followed beyond the point of star formation), though physical winding of the field is also lost.}
\label{fig:explodingstars}
\end{center}
\end{figure}

 A related but not identical issue, particularly for 3D calculations, is enforcement of the divergence-free constraint. Already what one means by ``divergence-free'' in SPH is unclear, because the stencil for measuring divergence of a vector field depends not only on the operator chosen but also on the particle distribution which can be different from one timestep to the next. It is primarily for this reason that the divergence cleaning schemes such as those discussed in \citet{pm05} are not very successful for 3D problems. Far better are approaches that have the divergence constraint ``built-in''. In particular using the `Euler potentials' formulation, ${\bf B} = \nabla\alpha \times \nabla\beta$, is a very elegant approach for a Lagrangian code, not only because the divergence is zero by construction but also because the induction equation adopts the very simple form $d\alpha/dt = 0; d\beta/dt = 0$ corresponding to the advection of field lines by Lagrangian particles. Use of the Euler potentials has enabled us to study the role of magnetic fields in realistic 3D star formation problems \citep{pb07,pb08,pb09}, where the meaning of \emph{not} maintaining the divergence constraint is clear. Figure~\ref{fig:explodingstars} gives an example of this, showing the results of a spherical, rotating collapse calculation with an initially uniform magnetic field threading the initial cloud. Using a ``standard'' SPMHD approach \citep{pm05}, the protostar simply explodes due to numerical force errors caused by the extremely large gradients and associated non-zero divergence in the magnetic field (left panel). Using the Euler potentials one is able to follow the calculation stably to well beyond the point of star formation (right panel).
 
  However, the ``advection'' property of the Euler potentials also means that physical winding of the fields is lost once a one-to-one mapping of the field from the initial conditions ceases to exist. For this reason we have recently explored an alternative approach based on the vector potential, the hope being that similar maintenance of the divergence-free condition could be achieved without the loss of physical winding processes. However the induction equation for the vector potential takes a rather more complicated form:
 \begin{equation}
 \frac{d{\bf A}}{dt} = {\bf v} \times \nabla \times {\bf A} + ({\bf v}\cdot\nabla) {\bf A} -\eta{\bf J} + \nabla \phi,
 \end{equation}
where ${\bf v}$ is the velocity, $\eta$ is the resistivity, ${\bf J}$ is the current and $\phi$ is an arbitrary scalar corresponding to the freedom to choose a gauge. The second term is particularly nasty since it represents effectively trying to compute ``reverse advection'' on the particles --- the absence of an advective term being the primary advantage of Lagrangian schemes over their Eulerian counterparts. Nevertheless Axel Brandenburg suggested that choosing $\phi = {\bf A}\cdot{\bf v}$ as a gauge might solve these problems since the gradients are flipped onto the ${\bf v}$ instead of ${\bf A}$. Choosing such a gauge also means that the induction equation becomes Galilean invariant, which in turn implies that it can be used to derive an exactly-conservative SPMHD formulation based on the vector potential \emph{ab initio} using a variational principle.

The great hope was that such a vector potential-based formulation, with the divergence-constraint ``built-in'' as in the Euler potentials method, could provide an elegant method which was both conservative and stable, without any of the numerical problems of previous formulations. Alas, this turned out not to be the case --- in fact the equations so derived \citep[see][]{price10}, whilst elegant, turned out to be even \emph{more} unstable to the clumping instability than in the standard approach. Furthermore it was found that \emph{another}, independent instability occurs related to the unconstrained growth of unphysical vector potential components, most likely related to poor accuracy with respect to the ``reverse-advection''-type terms. The latter means that the vector potential cannot be used even in conjunction with a standard (stable but non-conservative) SPMHD force term. The somewhat painful conclusion was therefore that the vector potential is not a viable approach for SPMHD.

%%% Top level section head (remove "%" symbol)
%\subsection{}   %%% Second level section head (remove "%" symbol)
%\subsubsection{}   %%% Lowest level section head (remove "%" symbol)
%\section*{}    %%% Unnumbered top level section head (remove "%" symbol)
%\subsection*{}   %%% Unnumbered second level section head (remove "%" symbol)

\acknowledgements %%% Text of acknowledgements runs on after this command.
DJP acknowledges the support of a Monash Fellowship. Figures in this paper were produced using \textsc{splash} \citep{splashpaper}. Thanks to the referee, Giuseppe Lanzafame, for suggesting Figure~\ref{fig:reynolds}.

\bibliography{sph,starformation,turbulence,chfeder,mhd}

%%% THE BIBLIOGRAPHY
%%%
%%% CONSULT SECTION 3 OF "INSTRUCTIONS FOR AUTHORS" FOR HOW TO USE NATBIB.
%%% AUTHORS ARE ENCOURAGED TO USE EITHER THE "THEBIBLIOGRAPY" ENVIRONMENT
%%% BY UNCOMMENTING (DELETING THE "%" SYMBOL) THE COMMANDS BELOW, OR BY
%%% USING THE BIBTEX ENVIRONMENT. TO FIND OUT WHICH IS APPLICABLE TO YOUR
%%% CONTRIBUTION, CONSULT THE VOLUME EDITORS FOR YOUR PROCEEDINGS.
%%%

%\begin{thebibliography}{}
%\bibitem[]{}
%\bibitem[]{}
%\bibitem[]{}
%\bibitem[]{}
%\bibitem[]{}
%\bibitem[]{}
%\bibitem[]{}
%\bibitem[]{}
%\bibitem[]{}
%\bibitem[]{}
%\bibitem[]{}
%\bibitem[]{}
%\end{thebibliography}

\end{document}